\documentclass[english,twocolumn]{revtex4}
\usepackage[T1]{fontenc}
\usepackage[latin9]{inputenc}
\usepackage{graphicx}

\makeatletter

\makeatletter

\makeatletter
\makeatother
\makeatother

\makeatother

\usepackage{babel}

\begin{document}

\title{Conductivity of electronic liquid-crystalline mesophases}

\author{Rafael M. Fernandes$^{1,2,3}$, {J\"org} Schmalian$^{1}$ and Harry
Westfahl Jr.$^{2}$ }

\date{\today}

\affiliation{$(1)$ Ames Laboratory and Department of Physics and Astronomy, Iowa
State University, Ames, IA 50011}

\affiliation{$(2)$ {Laborat\'{o}rio} Nacional de Luz {S\'{\i}ncrotron}, Caixa
Postal 6192, 13083-970 Campinas, SP, Brazil}

\affiliation{$(3)$ Instituto de {F\'{\i}sica} {}``Gleb Wataghin'', Universidade
Estadual de Campinas, 13083-970 Campinas, SP, Brazil}

\begin{abstract}
We investigate the connection between the transport properties and
the thermodynamics of electronic systems with a tendency to form broken-symmetry
mesophases evocative of the physics of liquid crystals. Through a
hydrodynamic approach to the electronic transport in inhomogeneous
systems, we develop a perturbative expansion for the macroscopic conductivity
to study the transport of two-dimensional smectic and nematic phases.
At the fluctuation induced first order phase transition expected for
the smectic to isotropic transition, a jump in the macroscopic conductivity
is predicted, with a directional dependence that reflects the fluctuation
spectrum of the order parameter. When elastic fluctuation modes melt
the smectic phase into a nematic phase, the resultant nematic order
parameter is shown to be linearly proportional to the conductivity
anisotropy. We also outline qualitative comparisons with recent experimental
works on strongly correlated materials that show evidences of electronic
liquid-crystalline mesophases. 
\end{abstract}

\maketitle

\section{Introduction}

Recent investigations on strongly correlated electron systems suggest
the emergence of inhomogeneous charge-ordered phases reminiscent from
the smectic and nematic states commonly found in liquid crystals \citep{kivelson1998fae}.
Such an analogy makes reference to the broken symmetries of each phase:
while the nematic phase breaks only rotational symmetry, the smectic
state also breaks the translational invariance along one particular
direction. Such electronic mesophases can be envisaged as {}``fluctuating''
and static charge stripes, respectively. Numerous experimental works
provide evidences for the occurrence of smectic charge-ordering in
nickelates \citep{cheong1994cos,huecker2006rcs}, manganites \citep{mori1998pco}
and $\mathrm{LBCO}$ cuprates \citep{tranquada2004qme}, as well as
indications of the existence of nematic order in quantum Hall systems
\citep{lilly1999eas}, ruthenates \citep{borzi2007fnf}, $\mathrm{LSCO}$
\citep{ando2002era,lavrov2003nsc} and $\mathrm{YBCO}$ cuprates \citep{bonetti2004etu,hinkov2008elc}.

The order of these mesophases is directly manifested in anisotropies
of the electric transport as well as in the magnetic or charge response
of the system, similar to the anisotropic optical and hydrodynamic
properties of liquid crystals, such as birefringence and viscosity
\citep{degennes1974plc}. An interesting question is the explicit
connection between the order parameters of these inhomogeneous mesophases
and their transport properties. While it is natural to expect that
the resistivity anisotropy is directly linked to an order parameter
that breaks rotational or translational symmetry, it is less obvious
how to connect transport properties to spatial or temporal fluctuations
and correlations of the system.

In this paper, we investigate the transport properties of inhomogeneous
electronic mesophases in the hydrodynamic transport limit, where collisions
dominate. In this regime, electron transport can be understood as
a diffusive process with inhomogeneous, potentially time dependent
diffusion constant $D\left(\mathbf{x,}t\right)$. We use a coarse
grained description and assume that the dephasing length scale $l_{0}$,
beyond which diffusive transport sets in, \ is small compared to
the length scale $\xi$ on which the inhomogeneities of the electronic
mesophases vary. The condition $l_{0}\ll\xi$ is expected to be valid
close to finite temperature phase transitions. This is certainly true
close to second order transitions where $\xi$ diverges. However,
even if the transition is weakly first order, we still expect $\xi$
to be larger than the lattice constant $a$ while $l_{0}$ becomes
comparable to $a$ due to strong inelastic scattering close to the
transition.

As will be discussed in detail below, our formalism for the conductivity
of inhomogeneous mesophases has a close connection to the theory of
random resistor network (RRN) \citep{kirkpatrick1973pac}. Previously,
the classical RRN was applied to explain transport properties of composite
films \citep{hurvits1993qat}, manganites \citep{mayr2001rmp} and
silver chalcogenides \citep{parish2003nsm}. Moreover, numerical simulations
of correlated versions of this network were carried out in the contexts
of manganites \citep{bastiaansen1997cpa}, disordered electronic nematic
phases in cuprates \citep{carlson2006han} and finite temperature
Mott transitions \citep{papanikolaou:026408}. The present approach
is an analytical theory for transport in electronic mesophases valid
for inhomogeneities that are small in amplitude but well correlated
in space. It can also be considered as an analytic theory for resistor
networks with correlated local conductivities \citep{blackman1976tcd}.
Here, the spatial correlations between the microscopic resistors is
directly connected to the correlations of the order parameter describing
the mesophase, and not to an arbitrary distribution function \citep{weinrib1984lrc}.
In the formalism that we develop, not only the order parameter mean
value, but also its relevant fluctuations are explicit related to
the macroscopic d.c. conductivity through a perturbative expansion.

By applying this general hydrodynamic transport model to the static
charge striped phase, we show that while the conductivity measured
perpendicular to the stripes probes mainly the order parameter amplitude,
the conductivity measured parallel to them is particularly sensitive
to the fluctuation spectrum. Thus, from transport measurements, it
is in principle possible to obtain information about the microscopic
character of the anisotropic mesophase, such as the first moments
of the Boltzmann distribution function, for example. As a specific
realization of the thermodynamics of the electronic smectic state,
we consider the Coulomb frustrated Ising model, first introduced by
Emery and Kivelson in the context of doped Mott insulators that present
high-temperature superconductivity \citep{emery1993fep,low1994sim}.
We obtain an analytic expression for the d.c. conductivity by using
the self-consistent mean-field solution of the effective Ginzburg-Landau
Hamiltonian, called the Brazovskii Hamiltonian \citep{brazovskii1975pti}.
We then show that the fluctuation induced first-order transition from
the isotropic liquid to the smectic phase is manifested as an anisotropic
jump of the conductivity, whose sign brings information about the
fluctuation spectrum of the order parameter. Qualitative comparisons
to experimental data showing jumps in the resistivity of nickelates
are also outlined.

In our theory, the investigation of the transport properties of the
electronic nematics focuses on the role of thermally excited modes.
Thus, it is complementary to the analysis of quantum modes as discussed
in the recent literature \citep{oganesyan2001qtn}. Based on the work
of Toner and Nelson \citep{toner1981sca}, we also describe the high-temperature
nematic phase as a smectic phase melted due to the elastic fluctuations
of stripes. We show explicitly that there is a temperature range where
the macroscopic conductivity anisotropy is linearly proportional to
the nematic order parameter, as expected from symmetry considerations
\citep{fradkin2000npt}, and determine the prefactor that connects
transport properties and nematic order. The same linear relation,
including the prefactor, was also observed in numerical simulations
of a \emph{disordered} electronic nematic phase at zero temperature
\citep{carlson2006han}, suggesting a close connection between the
two rather different approaches. Finally, a very recent experiment
regarding $\mathrm{YBCO}$ showed a remarkable resemblance between
the spectral weight of the low-energy anisotropic spin fluctuations,
obtained through neutron scattering, and the resistivity anisotropy,
obtained from transport measurements \citep{hinkov2008elc}.

The remainder of the paper is organized as follows: in section \ref{sec_diffusion_model},
we provide a detailed derivation of the hydrodynamic model for the
diffusion of an electron in an inhomogeneous medium, which will be
used throughout the work. Section \ref{sec_smectic} is devoted to
the application of this formalism to the transport properties of an
electronic smectic phase. Not only do we outline very general properties,
but we also obtain specific results after describing the charge stripes
thermodynamics by the Brazovskii model. In section \ref{sec_nematic},
the d.c. conductivity of an electronic phase with nematic type of
order is investigated through the hydrodynamic transport model. Such
a state is described as a smectic phase melted by thermally excited
elastic fluctuation modes of the stripes. Comparisons to other approaches
as well as to recent experiments involving doped transition metal
oxides are delineated. Finally, section \ref{sec_conclusions} is
devoted to the final remarks and acknowledgments.

\section{Diffusive transport in inhomogeneous mesophases \label{sec_diffusion_model}}

Let us consider the diffusion of an electron in an arbitrary inhomogeneous
medium. We start from the continuity equation \begin{equation}
\frac{\partial\rho\left(\mathbf{x},t\right)}{\partial t}+\mathbf{\nabla}\cdot\mathbf{j}\left(\mathbf{x},t\right)=0\:,\label{continuity}\end{equation}
 connecting the charge and current densities. The macroscopic conductivity
of the medium $\sigma_{\alpha\beta}$ relates the mean current density
to the external electric field $-\mathbf{\nabla}U_{\mathrm{ext}}\left(\mathbf{x,}t\right)$\begin{equation}
\left\langle j_{\alpha}\left(\mathbf{x},t\right)\right\rangle =-\int\sigma_{\alpha\beta}\left(t-t^{\prime}\right)\mathbf{\nabla_{\beta}}U_{\mathrm{ext}}\left(\mathbf{x,}t^{\prime}\right)dt^{\prime}\:,\label{aux_Ohm_law}\end{equation}
 yielding, in the Fourier space \begin{equation}
\left\langle \rho\left(\mathbf{k},\omega\right)\right\rangle =-i\frac{\sigma_{\alpha\beta}\left(\omega\right)k_{\alpha}k_{\beta}}{\omega}U_{\mathrm{ext}}\left(\mathbf{k},\omega\right)\:.\label{Ohm_law}\end{equation}

Next, we assume that locally the diffusive relation \begin{equation}
j_{\alpha}\left(\mathbf{x},t\right)=-\chi_{0}D_{\alpha\beta}\left(\mathbf{x,}t\right)\mathbf{\nabla_{\beta}}U_{\mathrm{loc}}\left(\mathbf{x,}t\right)\label{current_potential}\end{equation}
 holds for the current density in terms of the local electrical field
$-\mathbf{\nabla}U_{\mathrm{loc}}\left(\mathbf{x,}t\right)$, where
$D_{\alpha\beta}\left(\mathbf{x,}t\right)$ denotes the inhomogeneous
and possible time dependent local diffusion coefficient. Within linear
response, the uniform charge susceptibility $\chi_{0}$ connects the
local charge variations with the difference between the local and
external potentials \begin{equation}
\rho\left(\mathbf{x},t\right)=\chi_{0}\left[U_{\mathrm{loc}}\left(\mathbf{x,}t\right)-U_{\mathrm{ext}}\left(\mathbf{x,}t\right)\right]\:.\label{charge_susceptibility}\end{equation}

Combining the last two expressions, we are able to relate the external
and local potentials via \begin{equation}
\hat{G}^{-1}U_{\mathrm{loc}}\left(\mathbf{x,}t\right)=\frac{\partial U_{\mathrm{ext}}\left(\mathbf{x,}t\right)}{\partial t}\:,\label{local_external_potential}\end{equation}
 with inverse diffusion operator \begin{equation}
\hat{G}^{-1}=\frac{\partial}{\partial t}-\mathbf{\nabla_{\alpha}}\left[D_{\alpha\beta}\left(\mathbf{x,}t\right)\mathbf{\nabla_{\beta}}\right]\:.\label{differential_operator_G}\end{equation}

This yields, for the charge density \begin{equation}
\rho\left(\mathbf{x},t\right)=\chi_{0}\left[\hat{G}\frac{\partial}{\partial t}-1\right]U_{\mathrm{ext}}\left(\mathbf{x,}t\right)\:.\label{microscopic_ohm}\end{equation}

After taking the configurational average over $\rho\left(\mathbf{x},t\right)$,
we can Fourier transform its mean value and obtain, by comparing to
Eq. \ref{Ohm_law} \begin{equation}
\sigma_{\alpha\beta}\left(\omega\right)=\lim_{k\rightarrow0}\frac{i\omega}{k_{\alpha}k_{\beta}}\chi_{0}\left[-i\omega G\left(\mathbf{k},\omega\right)-1\right]\:,\label{ac_conductivity}\end{equation}
 where $G\left(\mathbf{k},\omega\right)\ $is the Fourier transform
of the average of the differential operator (\ref{differential_operator_G}).
In the d.c. limit, this finally yields \begin{equation}
\sigma_{\alpha\beta}=\chi_{0}\lim_{\omega\rightarrow0}\lim_{k\rightarrow0}\frac{\omega^{2}}{k_{\alpha}k_{\beta}}\mathrm{Re}\left[G\left(\mathbf{k},\omega\right)\right]\:.\ \label{dc_conductivity}\end{equation}

Using the Einstein relation \citep{chaikin1995pcm}, we can identify
the anisotropic macroscopic diffusion coefficient as $\lim\limits _{\omega\rightarrow0}\lim\limits _{k\rightarrow0}\frac{\omega^{2}}{k_{\alpha}k_{\beta}}\mathrm{Re}\left[G\left(\mathbf{k},\omega\right)\right]$.
Therefore, it is clear that, even if the tensor of the local diffusion
coefficient behaves as $D_{\alpha\beta}\left(\mathbf{x},t\right)=D\left(\mathbf{x,}t\right)\delta_{\alpha\beta}$,
the global diffusion coefficient can be anisotropic, as long as $D\left(\mathbf{x,}t\right)$
is a function of the direction of $\mathbf{x}$.

This hydrodynamic formalism has a one to one analogy with the theory
of random resistor networks (RRN) \citep{stephen1978mft}. In its
most elementary form, one considers two resistors with conductivities
$\sigma_{A}$ and $\sigma_{B}$, randomly distributed with probabilities
$p$ and $1-p$ over the links of a network. An external electric
potential $U_{\mathrm{ext}}$ is then applied in each site through
a local capacitor with specific capacitance $C$. As shown by \citet{stephen1978mft},
the macroscopic conductivity is given by a result identical to Eq.\ref{dc_conductivity},
with $C$ playing the role of the charge susceptibility while the
local conductivities of the RRN correspond to $\chi_{0}D\left(\mathbf{x}\right)$.
In the RRN problem, the average is performed over the distribution
function of the resistor network, which is frequently assumed to be
a binomial distribution, characterized by the probability $p$.

The key difference between our approach and the RRN theory is that
the distribution function for the local diffusion coefficient is determined
by the distribution function of the order parameter. Let the order
of an inhomogeneous electronic nematic or smectic state be characterized
by a scalar density field $\rho\left(\mathbf{x},t\right)$. Here,
$\rho\left(\mathbf{x},t\right)$ is the deviation of the the coarse
grained electron density from its mean value. We then assume a simple
connection\begin{equation}
D_{\alpha\beta}\left(\mathbf{x,}t\right)\equiv D_{\alpha\beta}\left[\rho\left(\mathbf{x,}t\right)\right]\label{formal_local_diffusion}\end{equation}
between the spatially varying diffusion coefficient and the electron
density of the electronic mesophase, meaning that the temporal and
spatial variations of the diffusion coefficient are determined solely
by those of $\rho\left(\mathbf{x,}t\right)$. In our cases of interest,
the ordered inhomogeneous state is characterized by an order parameter
that varies in space, alternating between $\rho>0$ and $\rho<0$.
For instance, in electronic smectics, $\rho>0$ ($\rho<0$) denotes
a hole rich (poor) coarse-grained region. Since each of these regions
has its own conducting properties, we can associate different local
conductivities to each of them.

For weakly inhomogeneous systems, we expand $D_{\alpha\beta}\left(\mathbf{x,}t\right)$
relative to the homogeneous state, where $\rho\left(\mathbf{x},t\right)=0$.
Hence, we propose the following specific form for $D_{\alpha\beta}\left(\mathbf{x,}t\right)$
of Eq.(\ref{formal_local_diffusion})

\begin{equation}
D_{\alpha\beta}\left(\mathbf{x,}t\right)=\chi_{0}\sigma_{0}\left[1+g\rho\left(\mathbf{x,}t\right)\right]\delta_{\alpha\beta}\:,\label{microscopic_cond}\end{equation}
where $\sigma_{0}$ is the microscopic conductivity mean value and
$g$ is the coupling constant measuring the contrast between the conductivities
of distinct regions. The physical meaning of these two parameters
can be better visualized considering the limit of an ordered \emph{homogeneous}
phase where $\rho=\pm\rho_{0}$. Denoting the conductivity of the
saturated homogeneous $\rho>0$ ($\rho<0$) phase as $\sigma_{>}$
($\sigma_{<}$), we have $\sigma_{0}=\left(\sigma_{>}+\sigma_{<}\right)/2$
and $g=\rho_{0}^{-1}\left(\sigma_{>}-\sigma_{<}\right)/\left(\sigma_{>}+\sigma_{<}\right)$.
A relation similar to Eq. \ref{microscopic_cond} was considered in
the context of the Mott transition by Papanikolaou et al. \citep{papanikolaou:026408}.
However, in that case it was necessary to include a next order term
$\propto\rho\left(\mathbf{x,}t\right)^{2}$ in the expansion, Eq.\ref{microscopic_cond},
originated from the contribution of interface scattering. Such effects
are not included in the present work.

Now, we particularize our analysis to the situation where the inhomogeneities
of the system are small, such that $g\ll1$. Thus, the diffusion differential
operator (\ref{differential_operator_G}) can be perturbatively expanded
as $\hat{G}={\displaystyle \sum_{n=0}^{\infty}}g^{n}\left(\hat{G}_{0}\hat{V}\right)^{n}\hat{G}_{0}$,
with

\begin{eqnarray}
\left(\hat{G}_{0}\right)^{-1} & = & \frac{\partial}{\partial t}-D_{0}\nabla^{2}\nonumber \\
\hat{V} & = & D_{0}\mathbf{\nabla}\cdot\left[\rho\left(\mathbf{x,}t\right)\mathbf{\nabla}\right]\:,\label{decomposition_G0_V}\end{eqnarray}
 where the bare diffusion coefficient is $D_{0}=\chi_{0}\sigma_{0}$.
In order to obtain a perturbative expansion for the conductivity through
Eq. \ref{dc_conductivity}, we need to perform an average over $\rho$
before taking the Fourier transform of $\hat{V}$. There are two opposite
limits one can consider: in the first one, which we shall call \emph{quenched}
limit, the electron diffuses faster than the field fluctuates, probing
a frozen configuration of the order parameter. Hence, after taking
the proper Fourier transforms, we obtain, to second order in $g$

\begin{eqnarray}
\sigma_{\alpha\alpha} & = & \sigma_{0}\left[1+g\left\langle \rho\left(\mathbf{k=0,}\omega=0\right)\right\rangle -\phantom{\frac{q^{2}\left(\hat{k}\cdot\vec{q}\right)^{2}}{\left(\nu^{2}C^{2}+\sigma_{0}^{2}q^{4}\right)}}\right.\label{aux_macroscopic_final}\\
 &  & \left.g^{2}\int d^{d}kd\omega\frac{k^{4}\left(\hat{\mathbf{n}}_{\alpha}\cdot\hat{\mathbf{k}}\right)^{2}}{\left(\omega^{2}D_{0}^{-2}+k^{4}\right)}\left\langle \rho\left(\mathbf{k,}\omega\right)\rho\left(-\mathbf{k,}-\omega\right)\right\rangle \right]\:,\nonumber \end{eqnarray}
where $\hat{n}_{\alpha}$ is the direction taken to measure the conductivity
and $\left\langle \cdots\right\rangle $ denotes the proper average
over the order parameter, which can be the usual thermodynamic average,
for example. Similarly we could also consider transport in nonequilibrium
configurations such as glassy states; then, $\left\langle \cdots\right\rangle $
refers to the corresponding dynamic average of the nonequilibrium
configuration under consideration \citep{schmalian2000sgs,westfahljr2001sgr}.
For future reference, we rewrite the previous formula for the special
situation where the field is static

\begin{eqnarray}
\sigma_{\alpha\alpha} & = & \sigma_{0}\left[1+g\left\langle \rho\left(\mathbf{k=0}\right)\right\rangle \right.\label{macroscopic_final_quenched}\\
 &  & \left.-g^{2}\int d^{d}k\left(\hat{\mathbf{n}}_{\alpha}\cdot\hat{\mathbf{k}}\right)^{2}\left\langle \rho\left(\mathbf{k}\right)\rho\left(-\mathbf{k}\right)\right\rangle \right]\:.\nonumber \end{eqnarray}

In the opposite limit, which we shall call \emph{annealed} limit,
we consider a field that fluctuates much faster than the electron
diffuses. Therefore, it is legitimate to replace the order parameter
by its mean value $\rho\rightarrow\left\langle \rho\right\rangle $
in the microscopic conductivity expression (\ref{microscopic_cond}),
yielding

\begin{eqnarray}
\sigma_{\alpha\alpha} & = & \sigma_{0}\left[1+g\left\langle \rho\left(\mathbf{k=0}\right)\right\rangle \right.\label{macroscopic_final_annealed}\\
 &  & \left.-g^{2}\int d^{d}k\left(\hat{\mathbf{n}}_{\alpha}\cdot\hat{\mathbf{k}}\right)^{2}\left\langle \rho\left(\mathbf{k}\right)\right\rangle \left\langle \rho\left(-\mathbf{k}\right)\right\rangle \right]\:.\nonumber \end{eqnarray}

It is clear that the anisotropic character of the conductivity is
not manifested until second order in $g$. Moreover, this second order
term reflects a fundamental difference between the quenched and annealed
limits. While in the latter the conductivity behavior is dictated
by the order parameter amplitude, in the former we note a tight connection
between the macroscopic conductivity $\sigma_{\alpha\alpha}$ and
the correlation function $\left\langle \rho\left(\mathbf{k}\right)\rho\left(-\mathbf{k}\right)\right\rangle $,
resembling other similar relations found in condensed matter systems,
such as between the scattering cross section and the thermodynamic
correlation functions.

\section{Transport in the smectic phase \label{sec_smectic}}

Using the formalism developed in the previous section, there are some
very general statements we can make about the conductivity of an electronic
smectic mesophase, independent on the specific model under consideration.
Hereafter, we consider that the static order parameter $\rho(\mathbf{x})$
describes local fluctuations of the charge density and that the electron
diffuses much faster than the field changes (quenched limit). Since
the system is usually electrically neutral, such that $\left\langle \rho\left(\mathbf{k=0}\right)\right\rangle =\frac{1}{V}\int\left\langle \rho(\mathbf{x})\right\rangle d^{d}x=0$,
the lowest non-vanishing correction to the uniform conductivity in
perturbation theory is of second order in the contrast. After writing
the correlation function as

\begin{equation}
\left\langle \rho\left(\mathbf{k}\right)\rho\left(-\mathbf{k}\right)\right\rangle =\mathcal{C}\left(\mathbf{k}\right)+\left\langle \rho\left(\mathbf{k}\right)\right\rangle \left\langle \rho\left(-\mathbf{k}\right)\right\rangle \:,\label{correlation_function}\end{equation}
where $\mathcal{C}\left(\mathbf{k}\right)$ is the connected correlation
function, we can split the second order term of the macroscopic conductivity
(\ref{macroscopic_final_quenched}) in two parts, $\sigma_{\alpha\alpha}=\sigma_{0}\left[1-g^{2}\left(\sigma_{\alpha\alpha}^{\prime}+\sigma_{\alpha\alpha}^{\prime\prime}\right)\right]$:

\begin{eqnarray}
\sigma_{\alpha\alpha}^{\prime} & = & \int d^{d}k\left(\hat{\mathbf{n}}_{\alpha}\cdot\hat{\mathbf{k}}\right)^{2}\mathcal{C}\left(\mathbf{k}\right)\nonumber \\
\sigma_{\alpha\alpha}^{\prime\prime} & = & \int d^{d}k\left(\hat{\mathbf{n}}_{\alpha}\cdot\hat{\mathbf{k}}\right)^{2}\left\langle \rho\left(\mathbf{k}\right)\right\rangle \left\langle \rho\left(-\mathbf{k}\right)\right\rangle \:.\label{conduct_two_parts}\end{eqnarray}

In the case of interest here, the correlation function depends only
on the modulus of the momentum, i.e. $\mathcal{C}\left(\mathbf{k}\right)=\mathcal{C}\left(k\right)$.
Therefore, $\sigma_{\alpha\alpha}^{\prime}$ does not depend on the
direction taken to measure the conductivity, since $\hat{\mathbf{n}}_{\alpha}$
will be integrated out over all directions. Hence, $\sigma_{\alpha\alpha}^{\prime}$
is isotropic and depends only on the order parameter fluctuation spectrum

\begin{equation}
\sigma_{\alpha\alpha}^{\prime}\propto\int d^{d}k\mathcal{C}\left(k\right)=\mathcal{C}\left(\mathbf{x},\mathbf{x}\right)=\left\langle \rho^{2}\left(\mathbf{x}\right)\right\rangle -\left\langle \rho\left(\mathbf{x}\right)\right\rangle ^{2}\:.\label{aux_relation_prime}\end{equation}

Meanwhile, the term $\sigma_{\alpha\alpha}^{\prime\prime}$ is proportional
to the averaged order parameter. For small amplitude static stripes,
it is reasonable to assume that the order parameter mean value is
strongly anisotropic and has a pronounced peak along the modulation
direction $\mathbf{q_{0}}$, such that $\left\langle \rho\left(\mathbf{k}\right)\right\rangle \sim A\delta\left(\mathbf{k}-\mathbf{q_{0}}\right)$.
Higher harmonics (entering as $\delta\left(\mathbf{k}-2\mathbf{q_{0}}\right)$
or $\delta\left(\mathbf{k}-3\mathbf{q_{0}}\right)$ etc.) only matter
once the amplitude of the inhomogeneity becomes large, i.e. $\rho\left(\mathbf{x,}t\right)$
becomes large compared to the mean electron density. It follows that,
unlike $\sigma_{\alpha\alpha}^{\prime}$, the term $\sigma_{\alpha\alpha}^{\prime\prime}$
is anisotropic and depends on the relative angle $\theta_{\alpha}$
between the external electric field and the modulation vector through

\begin{equation}
\sigma_{\alpha\alpha}^{\prime\prime}\propto A^{2}\cos^{2}\theta_{\alpha}\:.\label{aux_relation_prime_2}\end{equation}

Thus, if the conductivity is measured perpendicular to the stripes
$\left(\hat{\mathbf{n}}_{\alpha}\parallel\mathbf{q}_{\mathbf{0}}\right)$,
the order parameter and its fluctuation spectrum will be probed. However,
if the current is applied parallel to the stripes $\left(\hat{\mathbf{n}}_{\alpha}\perp\mathbf{q}_{\mathbf{0}}\right)$,
the measurement will be sensitive solely to the fluctuations of the
order parameter. Clearly, the transport quantity that only probes
the order parameter amplitude is the anisotropic conductivity $\sigma_{\perp}-\sigma_{\parallel}$.

Another interesting situation to investigate is when the system undergoes
a phase transition to the state with no broken symmetries, the isotropic
liquid. For instance, let us consider a first order transition from
the liquid to the smectic phase: if the conductivity is measured parallel
to the stripes direction, it is expected that $\sigma_{\alpha\alpha}^{\prime\prime}$
vanishes, and the isotropic part $\sigma_{\alpha\alpha}^{\prime}$
will dictate the behavior of the d.c. conductivity. Since fluctuations
of the order parameter are usually larger in the liquid disordered
state, we expect from Eq. \ref{aux_relation_prime} that, close to
the transition, the conductivity of the liquid phase will be smaller
than the one of the ordered smectic phase, $\sigma_{\parallel}^{liq}<\sigma_{\parallel}^{smc}$.

Instead, if the conductivity is measured perpendicular to the stripes
direction, the contribution of the term $\sigma_{\alpha\alpha}^{\prime\prime}$
will be maximum. Unlike its counterpart $\sigma_{\alpha\alpha}^{\prime}$,
this term is greater in the ordered side, since the order parameter
amplitude vanishes in the liquid phase. Hence, the question of whether
$\sigma_{\perp}^{dis}$ or $\sigma_{\perp}^{smc}$ is larger close
to the phase transition depends on the ratio between the jumps of
the fluctuations and of the amplitude of the order parameter. For
strong first order transitions, we expect the latter to be more significant,
meaning that $\sigma_{\perp}^{liq}>\sigma_{\perp}^{smc}$.

Let us now apply the hydrodynamic transport formalism to a specific
model for the thermodynamics of the electronic smectics. In particular,
we focus on doped layered transition metal oxides, which are believed
to be well described as $2D$ doped Mott insulators \citep{emery1999sph}.
In these compounds, the stripes appear as a compromise between a short-range
tendency of the doped charges to phase separate from the antiferromagnetic
background spins and the long-range electrostatic repulsion between
alike charges \citep{emery1993fep}. A simple model that attempts
to capture these properties is the Coulomb-frustrated Ising model
\citep{low1994sim}

\begin{eqnarray}
\mathcal{H}_{\mathrm{Ising-Coulomb}} & = & \frac{1}{2}\int d^{2}x\left\{ \tau_{0}\rho^{2}+\left\vert \nabla\rho\right\vert ^{2}+\frac{u}{4}\rho^{4}\right\} \nonumber \\
 &  & +\frac{q_{0}^{3}}{2 \pi} \int d^{2}x\frac{\rho(\mathbf{x})\rho(\mathbf{x}^{\prime})}{\left\vert \mathbf{x}-\mathbf{x}^{\prime}\right\vert }\:,\label{Coulomb_frustrated_Ising}\end{eqnarray}
where $\rho(\mathbf{x})>0$ denotes a hole-rich region (larger conductivity)
while $\rho(\mathbf{x})<0$ denotes a hole-poor region (smaller conductivity),
with $\int\left\langle \rho(\mathbf{x})\right\rangle d^{2}x=0$. As
usual, $\tau_{0}$ denotes the reduced temperature and $u$ is an
effective parameter. While the first term of the above Ginzburg-Landau
Hamiltonian expresses the tendency of phase separation, the second
term accounts for the frustration introduced by the Coulomb interaction.
This Hamiltonian is minimized in the Fourier space for a non-zero
wave vector whose modulus is $q_{0}$. Thus, considering the low-energy
fluctuation modes, we can expand Eq. \ref{Coulomb_frustrated_Ising}
around its minimum to obtain the effective functional:

\begin{eqnarray}
\mathcal{H} & = & \frac{1}{2}\int d^{2}xd^{2}x^{\prime}\rho(\mathbf{x})\mathcal{C}_{0}^{-1}\left(\mathbf{x},\mathbf{x}^{\prime}\right)\rho(\mathbf{x})+\frac{u}{4}\int d^{2}x\rho^{4}(\mathbf{x})\nonumber \\
 &  & \mathcal{C}_{0}^{-1}\left(\mathbf{x},\mathbf{x}^{\prime}\right)=\frac{1}{\left(2\pi\right)^{2}}\int d^{2}k\frac{\mathrm{e}^{i\mathbf{k}\cdot(\mathbf{x}-\mathbf{x}^{\prime})}}{\tau_{0}+\left(k-q_{0}\right)^{2}}\:.\label{brazo_Hamiltonian}\end{eqnarray}

Note that although $\tau_{0}$ in Eq. \ref{brazo_Hamiltonian} is shifted with 
respect to $\tau_{0}$ in Eq. \ref{Coulomb_frustrated_Ising}, 
we keep the same symbol to simplify the notation. 
This model was first studied by Brazovskii in the context of cholesteric
liquid crystals \citep{brazovskii1975pti}, and has been employed
to describe a variety of other physical systems with inhomogeneous
states, from {}``hard matter'' ones, such as dipolar ferromagnets
\citep{garel1982pts,fernandes:144421}, to {}``soft matter'' systems
like diblock copolymers \citep{fredrickson:697} and microemulsions
\citep{wu2002tmg}. The striped phase with modulation vector $\mathbf{q_{0}}$,
which has smectic order, is a thermodynamic stable state of the model.
In two dimensions, extra external potentials are necessary in order
to ensure the stability of this smectic phase against elastic fluctuations
- like crystalline fields \citep{abanov1995pdu} or pinning centers,
for example.

\begin{figure}

\begin{centering}
\includegraphics[width=0.9\columnwidth]{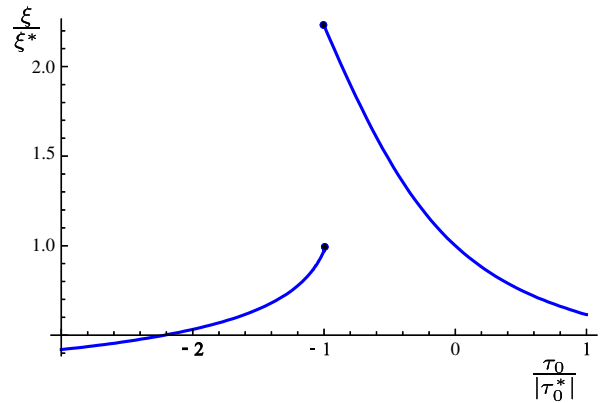} 
\par\end{centering}

\caption{Correlation length $\xi$ (in units of the correlation length $\xi^{*}$
of the smectic phase at the reduced transition temperature $\tau_{0}^{*}$)
as function of the reduced temperature $\tau_{0}$. }

\label{fig_jump} 
\end{figure}

Since we are interested in the main qualitative transport properties
of the system, we use the self-consistent mean-field solution of Eq.
\ref{brazo_Hamiltonian}. The details of this method are presented
elsewhere \citep{brazovskii1975pti,hohenberg1995mfd,westfahljr2005cdr,fernandes:144421}.
It predicts a fluctuation induced first order transition from the
smectic state $\left\langle \rho(\mathbf{x})\right\rangle =2A\cos\left(\mathbf{q_{0}}\cdot\mathbf{x}\right)$
to the liquid disordered phase. The temperature for which their free
energies become equal is $\tau_{0}^{\ast}\approx-\left(3\sqrt{2}q_{0}u\right)^{2/3}$,
which we will consider the transition temperature hereafter (this
is valid for an adiabatic change of the temperature). At this point,
there is a jump in the correlation length of the system $\xi$, as
sketched in figure \ref{fig_jump}. We consider the stripes frozen
in the diffusive electron reference frame; hence, substituting the
self-consistent solution of (\ref{brazo_Hamiltonian}) in Eq. \ref{macroscopic_final_quenched}
yields, for the conductivity of the liquid disordered and smectic
phases

\begin{figure}

\begin{centering}
\includegraphics[width=0.9\columnwidth]{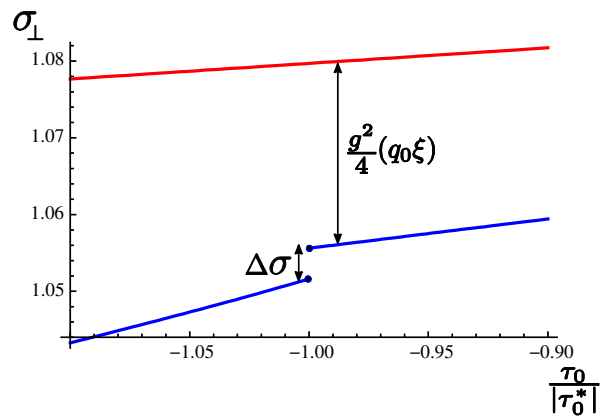} 
\par\end{centering}

\caption{Conductivity $\sigma_{\perp}$ measured perpendicular ($\theta_{\alpha}=0$)
to the stripes direction (solid line) as function of the reduced temperature
$\tau_{0}$ (in units of the transition temperature modulus $\left|\tau_{0}^{*}\right|$).
An intrinsic insulating behavior was considered for $\sigma_{0}$
(dashed line). }

\label{fig_jump_conductivity} 
\end{figure}

\begin{eqnarray}
\sigma_{\alpha\alpha}^{liq} & \simeq & \sigma_{0}\left\{ 1-\frac{g^{2}}{4}\left(q_{0}\xi\right)\right\}  
\label{conduct_striped} \\
\sigma_{\alpha\alpha}^{smc} & \simeq & \sigma_{0}\left\{ 1-\frac{g^{2}}{4}\left(q_{0}\xi\right)\left[1+4\left(\frac{\xi^{*}}{\xi^{\phantom{*}}}\right)^{3}\cos^{2}\theta_{\alpha}\right]\right\} \: , \nonumber
\end{eqnarray}
where $\xi$ is the correlation length, $\xi^{*}$ is its value at
the transition temperature for the smectic phase and $\theta_{\alpha}$
is the angle between the direction taken to measure the conductivity
and the direction of the stripes modulation, $\mathbf{q_{0}}$. As
expected, the conductivity is intrinsically anisotropic and it is
greater when measured parallel to the stripes $(\theta_{\alpha}=\frac{\pi}{2})$
than when it is measured perpendicular to them $(\theta_{\alpha}=0)$,
$\sigma_{\parallel}^{smc}>\sigma_{\perp}^{smc}$. Moreover, this anisotropy
increases as the temperature is lowered due to the decrease in $\xi$.
Another manifestation of the thermodynamic behavior in the transport
properties is the occurrence of a jump in the conductivity at the
transition temperature, as shown in figure \ref{fig_jump_conductivity}.

\begin{figure}

\begin{centering}
\includegraphics[width=0.9\columnwidth]{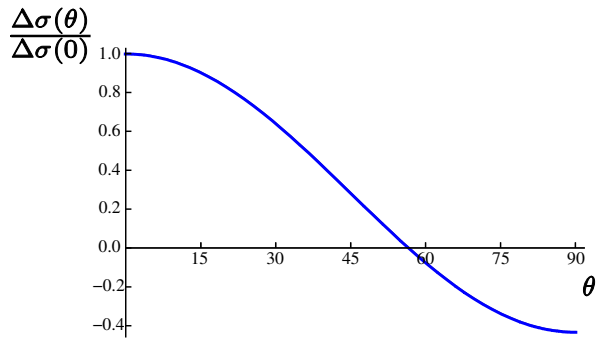} 
\par\end{centering}

\caption{Conductivity jump $\Delta\sigma\left(\theta\right)$ as function of
the angle $\theta$ between the modulation direction $\mathbf{q_{0}}$
and the direction taken to measure the conductivity (in degrees). }

\label{fig_sign} 
\end{figure}

Not only does the module of the jump $\Delta\sigma\left(\theta_{\alpha}\right)=\left.\sigma_{\alpha\alpha}^{liq}\right|_{\tau_{0}^{*}}-\left.\sigma_{\alpha\alpha}^{smc}\right|_{\tau_{0}^{*}}$
depend on the angle between the external current and the stripes modulation,
but also its sign. In figure \ref{fig_sign}, we plot $\Delta\sigma\left(\theta_{\alpha}\right)$
as a function of $\theta_{\alpha}$ and see that while its sign is
positive when the conductivity is measured perpendicular to the stripes,
it becomes negative when the measurement is made along the stripes.

This is in agreement with the general discussion carried out in the
beginning of the present section. The negative sign is a consequence
of the fact that the conductivity measured parallel to the stripes
$(\theta_{\alpha}=\frac{\pi}{2})$ probes the fluctuations of the
order parameter, which are greater in the disordered phase, as it
can be noted directly from the behavior of the correlation length
presented in figure \ref{fig_jump}. When the conductivity is measured
perpendicular to the stripes $(\theta_{\alpha}=0)$, even though fluctuations
still contribute to the conductivity, there is an extra contribution
coming from the order parameter amplitude, which is non-zero only
in the ordered phase. The positive sign is a balance between these
two contributions, meaning that the latter, proportional to $\xi^{-2}$,
is greater than the first, proportional to $\xi$.

In the LTO phase of the layered nickelates $\mathrm{Nd_{1-x}Sr_{x}NiO_{4}}$,
a sudden upturn in the resistivity was observed at the charge-ordering
temperature \citep{cheong1994cos,huecker2006rcs}. The external current
in the experiment of H\"ucker et al. \citep{huecker2006rcs} was
applied perpendicular to the direction of the stripes and this sudden
change could be the result of a jump rounded by disorder \citep{imry1979iqi}.
It would be interesting to realize an experiment where the conductivity
is measured along the stripes direction in order to check if an opposite
change in the resistivity takes place. This would provide additional
criteria to check the validity range of this hydrodynamic transport
model and send light on the fluctuation spectrum in a state of mesoscale
ordering.

In another class of nickelates, the $\mathrm{La_{2}NiO_{4+\delta}}$,
experiments probing resistivity fluctuations indicated no jump, but
also suggested a strong coupling to sample disorder, giving rise to
slow stripe dynamics \citep{pautrat:125106}. The current was applied
perpendicular to the stripes direction. For slow sample cooling, however,
jumps were observed below the transition temperature, inside the smectic
phase, with an increase in the resistivity. This was attributed to
changes in the modulation wave vector modulus $q_{0}$ due to the
coupling to the lattice, in accordance with neutron-diffraction measurements
that verified a decrease in $q_{0}$ \citep{wochner1998nds}. A possible
explanation is given in the context of the present model: considering
a reduction in the wave vector modulus $q_{0}^{\prime}=\eta q_{0}$,
$\eta<1$, and solving the resulting self-consistent Brazovskii equations,
we obtain $\sigma_{\perp}^{smc}\left(q_{0}^{\prime}\right)<\sigma_{\perp}^{smc}\left(q_{0}\right)$
for $\eta$ close to $1$ and for any temperature below the transition
one. Although other mechanisms could also explain this decrease in
the conductivity, like a temperature dependent contrast $g$, for
example, it is worth pointing that our model is consistent with this
experimental observation.

\section{Transport in the nematic phase \label{sec_nematic}}

In the previous section, we implicitly assumed the underlying presence
of crystalline fields and/or pinning disorder in the system, in order
to keep the smectic phase stable against low-energy fluctuation modes
of the stripes. In the absence of such stabilizing mechanisms, it
is well known that, at any finite temperature, the clean two-dimensional
smectic phase is unstable towards the formation of a nematic phase
due to elastic fluctuations of the stripes walls \citep{toner1981sca}.
Therefore, the nematic phase can be conceived as a melted smectic
phase, where the stripes have no longer true positional order. Since
two is the lower critical dimension of the resulting system, the $2D$
nematic state has only quasi-long range orientational order, which
is lost above the Kosterlitz-Thouless transition temperature $T_{KT}$
where pairs of disclinations unbind, driving the system to the isotropic
liquid phase.

Fluctuation modes of stripe walls other than thermally excited ones
can also melt the smectic phase. For example, random-field disorder
\citep{carlson2006han} and quantum fluctuations \citep{kivelson1998fae}
are able to drive the system to a nematic state even at zero temperature.
Alternatively, the quantum electronic nematics can be envisaged not
as a result of the quantum melting of the smectic phase, but as a
consequence of quadrupolar Pomeranchuk instabilities of the Fermi
surface of the isotropic liquid phase \citep{oganesyan2001qtn,sun:085124}.
Here, we will not consider the role of quantum fluctuations neither
the presence of disorder, but will follow Toner and Nelson's description
of the nematic phase, taking into account only thermally excited elastic
fluctuations \citep{toner1981sca}. This is fully consistent with
our hydrodynamic approach where the dephasing length is assumed short
as the system is considered at finite $T$.

Without loss of generality, we consider stripes modulated along the
$\hat{x}$ direction and substitute the order parameter $\rho(\mathbf{x})=2A\cos\left[q_{0}\left(x+u(x,y)\right)\right]$
in Eq. \ref{brazo_Hamiltonian}, where the displacement field $u(\mathbf{x})$
describes the elastic fluctuations. The resulting elastic Hamiltonian
is given, in the harmonic approximation, by

\begin{equation}
\mathcal{H}_{\mathrm{elastic}}=\frac{1}{2}B\int d^{2}x\left[\left(\frac{\partial u}{\partial x}\right)^{2}+\lambda^{2}\left(\frac{\partial^{2}u}{\partial y^{2}}\right)^{2}\right]\:,\label{elastic_hamiltonian}\end{equation}
 with $B=A^{2}q_{0}^{2}$ and $\lambda^{-1}=2q_{0}$. In two dimensions,
explicit calculations of the mean value and of the correlation function
of the order parameter yield $\left\langle \rho\right\rangle =0$
and

\begin{equation}
\left\langle \rho(\mathbf{x})\rho(0)\right\rangle \propto\left\{ \begin{array}{cc}
\exp\left[-\frac{q_{0}^{2}}{B}\left(\frac{\left\vert x\right\vert }{4\pi\lambda}\right)^{1/2}\right] & \:,\:\mathrm{for}\:\:\:\lambda x\gg y^{2}\\
\exp\left[-\frac{q_{0}^{2}}{4B\lambda}\left\vert y\right\vert \right] & \:,\:\mathrm{for}\:\:\:\lambda x\ll y^{2}\end{array}\right.\label{correlation_melted}\end{equation}
meaning that the system has not true positional order. After taking
into account the roles of two kinds of topological defects, dislocations
and disclinations, it follows that the elastic Hamiltonian describing
the melted state is the Frank Hamiltonian \citep{toner1981sca}

\begin{equation}
\mathcal{H}_{\mathrm{Frank}}=\frac{1}{2}\int d^{2}x\left\{ K_{1}\left(\mathbf{\nabla}\cdot\mathbf{N}\right)^{2}+K_{3}\left[\mathbf{N}\times\left(\mathbf{\nabla}\times\mathbf{N}\right)\right]^{2}\right\} \:,\label{Frank_Hamiltonian}\end{equation}
 which describes the elastic properties of a $2D$ nematic liquid
crystal. In the previous expression, the unit vector $\mathbf{N}$
denotes the nematic director and the Frank constants $K_{1}$ and
$K_{3}$ are functions of the elastic parameters $B$ and $\lambda$,
as well as of the energy necessary to excite an isolated dislocation,
$E_{D}$.

\begin{figure}
\begin{centering}
\includegraphics[width=0.9\columnwidth]{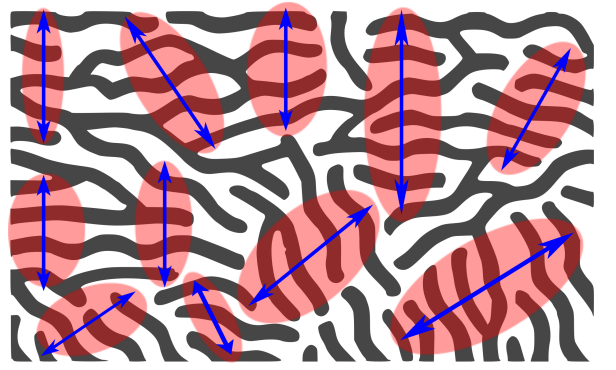} 
\par\end{centering}

\caption{Schematic representation of the nematic phase originated from the
melting of the smectic state due to thermally excited elastic fluctuations.
The blobs comprehend regions free of topological defects enclosing
stripes that are decorrelated by elastic fluctuations only. The director
of each blob is indicated by a double arrow. }

\label{fig_blobs} 
\end{figure}

The role of the topological defects is fundamental to the understanding
of the thermodynamics of the resulting nematic phase. As explained
above, the unbinding of disclination pairs induces a Kosterlitz-Thouless
transition to the isotropic liquid phase. Moreover, since isolated
dislocations have a finite excitation energy $E_{D}$, a new length
scale $\xi_{D}$ is introduced in the system below $T_{KT}$. Such
a length scale denotes the correlation length of isolated dislocations
that proliferate in the system as the temperature is increased, and
is given by $\xi_{D}=n_{D}^{-1/2}$, where $n_{D}=d^{-2}\exp\left(-\frac{E_{D}}{k_{B}T}\right)$
is the dislocations density and $d\propto q_{0}^{-1}$ is the stripes
mean width. For lengths greater than $\xi_{D}$ the system has the
same properties of a nematic liquid crystal, while for smaller lengths
there is smectic order decorrelated by elastic fluctuations only.
Therefore, one can consider the nematic phase as a set of finite size
smectic blobs \citep{toner1981sca}, where each blob has orientational
order given by the average of the orientation of the stripes enclosed,
as sketched in figure \ref{fig_blobs}.

We can now apply the formulae of the previous section to calculate
the conductivity of an individual blob $\sigma_{\alpha\alpha}^{blob}$.
Inside it, although density fluctuations remain small as long as $\tau_{0}<\tau_{0}^{\ast}$,
elastic fluctuations are relevant. While a diffusive electron takes
the typical time $t_{el}\sim\sigma_{0}^{-1}\xi_{D}^{2}$ to cross
the blob, phonon fluctuations propagate with a typical time scale
$t_{ph}\sim v^{-1}\xi_{D}$, where $v$ is a sound velocity. Therefore,
we have to substitute the order parameter mean value $\left\langle \rho\left(\mathbf{k}\right)\right\rangle =A\left[\delta\left(\mathbf{k}-\mathbf{q_{0}}\right)+\delta\left(\mathbf{k}+\mathbf{q_{0}}\right)\right]\mathrm{e}^{-W}$,
where $W=\mathrm{e}^{\frac{E_{D}}{3k_{B}T}}\left/\left(2^{\frac{2}{3}}\pi^{\frac{1}{3}}A^{2}\right)\right.$
is the Debye-Waller factor of the finite-size blob, in the annealed
limit of the macroscopic conductivity, Eq. \ref{macroscopic_final_annealed}.
We obtain

\begin{equation}
\sigma_{\alpha\alpha}^{blob}\simeq\sigma_{0}\left(1-2g^{2}A^{2}\mathrm{e}^{-2W}\cos^{2}\theta_{\alpha}\right)\:,\label{conduct_blob}\end{equation}
where $\theta_{\alpha}$ is the angle between the direction taken
to measure the conductivity and the blob director (we are following
the convention that the director is perpendicular to the stripes orientation
inside each blob). This procedure is valid for intermediate temperatures
only, where $W$ is small and the elastic fluctuations $\left\vert \delta\rho_{el}\right\vert =\sqrt{\mathrm{e}^{2W}-1}\left\langle \rho_{el}\right\rangle $
are not so large.

To obtain the system's macroscopic conductivity, the fluctuations
of the nematic director inside each blob have to be taken into account.
We can consider them slow in the electron reference frame. Thus, while
taking the annealed limit to calculate the conductivity inside each
blob, we have to take the quenched limit to calculate the conductivity
of the whole system, averaging over all blobs. We obtain

\begin{eqnarray}
\sigma_{xx}^{nem} & \simeq & \bar{\sigma}_{0}\left(1-\bar{g}\left\langle \cos2\zeta\right\rangle \right)\nonumber \\
\sigma_{yy}^{nem} & \simeq & \bar{\sigma}_{0}\left(1+\bar{g}\left\langle \cos2\zeta\right\rangle \right)\label{macroscopic_nematic}\end{eqnarray}
for the two opposite situations where the direction taken to measure
the conductivity is parallel ($\left\langle \theta_{\alpha}\right\rangle =0$,
$\hat{x}$ axis) or perpendicular ($\left\langle \theta_{\alpha}\right\rangle =\pi/2$,
$\hat{y}$ axis) to the mean orientation of the nematic director.
In the previous expression, $\zeta=\theta_{\alpha}-\left\langle \theta_{\alpha}\right\rangle $
denotes the director deviation from the mean value, $\bar{\sigma}_{0}=\sigma_{0}\left(1-g^{2}A^{2}\mathrm{e}^{-2W}\right)$
and $\bar{g}=g^{2}A^{2}\mathrm{e}^{-2W}$ (to second order in the
original $g$).

It is remarkable that the nematic order parameter $\left\langle \cos2\zeta\right\rangle $
appears spontaneously in the conductivity expression, even though
it was not assumed any coupling between the microscopic conductivity
and the nematic order parameter. Such a relation can be cast in a
more formal way by considering the tensorial nematic order parameter
in two dimensions, $\left\langle S_{\alpha\beta}\right\rangle =S\left(N_{\alpha}N_{\beta}-\frac{1}{2}\delta_{\alpha\beta}\right)$,
with $S=\left\langle 2\cos^{2}\zeta-1\right\rangle $. It is clear,
from Eq. \ref{macroscopic_nematic}, that the anisotropic conductivity
can be written as $\sigma_{\alpha\beta}^{nem}\simeq\bar{\sigma}_{0}\left(1-\bar{g}\left\langle S_{\alpha\beta}\right\rangle \right)$.

Another direct consequence of Eq. \ref{macroscopic_nematic} is that
the conductivity anisotropy is proportional to the nematic order parameter,
$\sigma_{yy}^{nem}-\sigma_{xx}^{nem}=2\sigma_{0}g^{2}A^{2}\mathrm{e}^{-2W}\left\langle \cos2\zeta\right\rangle $,
in agreement with symmetry arguments \citep{fradkin2000npt}. By comparing
this expression with the one referring to the conductivity anisotropy
of the smectic phase, Eq. \ref{conduct_striped}, we note that in
the present case the anisotropy is much smaller due to the Debye-Waller
factor $\mathrm{e}^{-2W}$. Moreover, this result agrees exactly with
the relation\begin{equation}
\left(\frac{r+1}{r-1}\right)\left(\frac{R_{\perp}-R_{\parallel}}{R_{\perp}+R_{\parallel}}\right)=\left\langle \cos2\zeta\right\rangle \label{relation_carlson}\end{equation}
found for $T=0$ by Carlson et al. \citep{carlson2006han}, with the
same prefactors (in our case, $r\equiv\frac{R_{\perp}(\left\langle \cos2\zeta\right\rangle \rightarrow1)}{R_{\parallel}(\left\langle \cos2\zeta\right\rangle \rightarrow1)}=\frac{1+\bar{g}}{1-\bar{g}}$).
In that work, the disordered electronic nematics was mapped on the
Random Field Ising Model (RFIM), and the conductivity was obtained
numerically through a random resistor network approach. Although the
two systems are very different, since ours is clean whereas the one
investigated by Carlson et al. is disordered, this alike outcome seems
to be a consequence of the existence of clusters with short-range
smectic order inside the nematic phase - the thermally excited blobs
in our model and the disorder generated nematic patches in the other
work.

Such a linear relation between a macroscopic quantity and the thermodynamic
order parameter can be very useful in the investigation of the orientational
ordering of nematic phases in the cuprates, specially through transport
measurements. In fact, a recent experiment \citep{hinkov2008elc}
on $\mathrm{YBa_{2}Cu_{3}O_{6.45}}$ suggested a linear relation between
the compound conductivity anisotropy and the spectral weight of the
low-energy anisotropic spin fluctuations. Still regarding possible
experimental implications, we also point that due to the large elastic
fluctuations intrinsic to the system's dimensionality, deviations
from the averaged conductivity are expected for small systems, what
would be manifested as noise in time series measurements.

Before finishing this section, an important remark regarding the mean
value $\left\langle \cos2\zeta\right\rangle $ should be made. Strictly
speaking, in two dimensions this average is zero in the thermodynamic
limit, due to Mermin-Wagner theorem (see, for instance, Chaikin and
Lubensky \citep{chaikin1995pcm}). However, in real layered systems,
finite-size effects as well as small interlayer couplings are able
to stabilize the nematic phase, granting a non-vanishing value for
the order parameter.

\section{concluding remarks \label{sec_conclusions}}

We presented a hydrodynamic transport theory that can provide important
tools in the investigation of electronic phases with smectic and nematic
symmetries, rendering explicit relations between macroscopic transport
quantities and the microscopic order parameter. Particularly, we considered
doped layered transition metal oxides, using the Brazovskii model
to describe the thermodynamics of the low-energy charge modes of their
smectic phases. The directional dependence of the sign of the conductivity
jump was shown to be a manifestation of the Brazovskii fluctuation
spectrum and a general characteristic of the hydrodynamic transport
model, constituting an interesting criterion to decide on the applicability
of the model to describe the observed static charge striped phases.

The finite temperature electronic nematics was conceived as a smectic
Brazovskii phase melted by thermally excited elastic fluctuations
of the stripes walls, following the approach of Toner and Nelson \citep{toner1981sca}.
Not only does the nematic order parameter appear explicitly and spontaneously
in our formalism, but it is also shown to be linearly proportional
to the conductivity anisotropy, following the same relation found
numerically by Carlson et al. \citep{carlson2006han} for the case
of the disordered electronic nematics at zero temperature.

In the context of nematic phases, it would be interesting to investigate
other excitations that are also able to melt the smectic phase in
a nematic state, such as quantum and disorder induced fluctuations \citep{kivelson1998fae,carlson2006han}.
Moreover, additional studies on out of equilibrium properties of the
electronic nematics, particularly on the connection between fluctuations
of the smectic clusters and transport time-series measurements, would
also be desirable to provide a richer picture of the problem. Finally,
applications of the general formalism to systems other than doped
transition metal oxides can also be envisaged, for example, for the
low-temperature striped phases that appear in quantum Hall systems
\citep{fradkin2000npt,oganesyan2001qtn,sun:085124} and for the inhomogeneous
phases found in spin glasses and Mott insulators \citep{papanikolaou:026408}.

The authors would like to thank C. Batista, V. Dobrosavljevic, E.
Fradkin, S. Papanikolaou, P. Phillipps, and P. G. Wolynes for helpful
discussions. This research was supported by CAPES and CNPq (Brazil)
and by the Ames Laboratory, operated for the US Department of Energy
by Iowa State University under Contract No. DE-AC 02-07CH11358. \bibliographystyle{apsrev}

\end{document}